%% file: ss_sr_nikkei.tex
\documentclass{svmult}

\usepackage{makeidx}         
\usepackage{graphicx}        
\usepackage{multicol}        
\usepackage[bottom]{footmisc}

\makeindex             

\begin{document}

\title*{Emergence of two-phase behavior in markets through
interaction and learning in agents with bounded rationality}
\titlerunning{Emergence of two-phase behavior in markets}
\author{Sitabhra Sinha\inst{1}\and
S. Raghavendra\inst{2}}
\institute{The Institute of Mathematical Sciences,
CIT Campus, Taramani, Chennai 600113, India.\\
\texttt{sitabhra@imsc.res.in}
\and Madras School of Economics, Anna University Campus, Chennai 600 025, 
India.}
%
%
\maketitle

Phenomena which involves collective choice of many agents who are interacting
with each other and choosing one of several alternatives, based on the
limited information available to them, frequently show switching between two
distinct phases characterized by a bimodal and an unimodal
distribution respectively. Examples include financial markets, movie
popularity and electoral behavior. Here we present a model for this biphasic
behavior and argue that it arises from interactions in a local neighborhood 
and adaptation \& learning based on information about the
effectiveness of past choices.

\section{Introduction}
\label{sec:1}
\vspace{-0.25cm}
The behavior of markets and other social agglomerations are made up of the
individual decisions of agents, choosing among a number of possibilities
open to them at a given time. Let us consider the example of binary choice,
where the agent can make one of two possible decisions, e.g., to buy or
to sell. If each agent makes a choice completely at random, the outcome
will be an unimodal distribution, a Gaussian to be precise, of the 
collective choice (i.e., the sum total of all the individual decisions),
at whose mean value the distribution will have its peak. 
In our example this implies that, on the average, there are equal numbers
of buyers and sellers.

However, empirical data in financial markets \cite{Ple03,Zhe04}, 
movie popularity
\cite{Sin04} and electoral behavior \cite{May74} indicate that
there is another
phase, corresponding to the agents predominantly choosing one
option over the other. This is reflected in a bimodal distribution of the
collective choice (Fig. 1).

\begin{figure}[htbp]
\centering
\includegraphics[width=0.48\linewidth,clip] {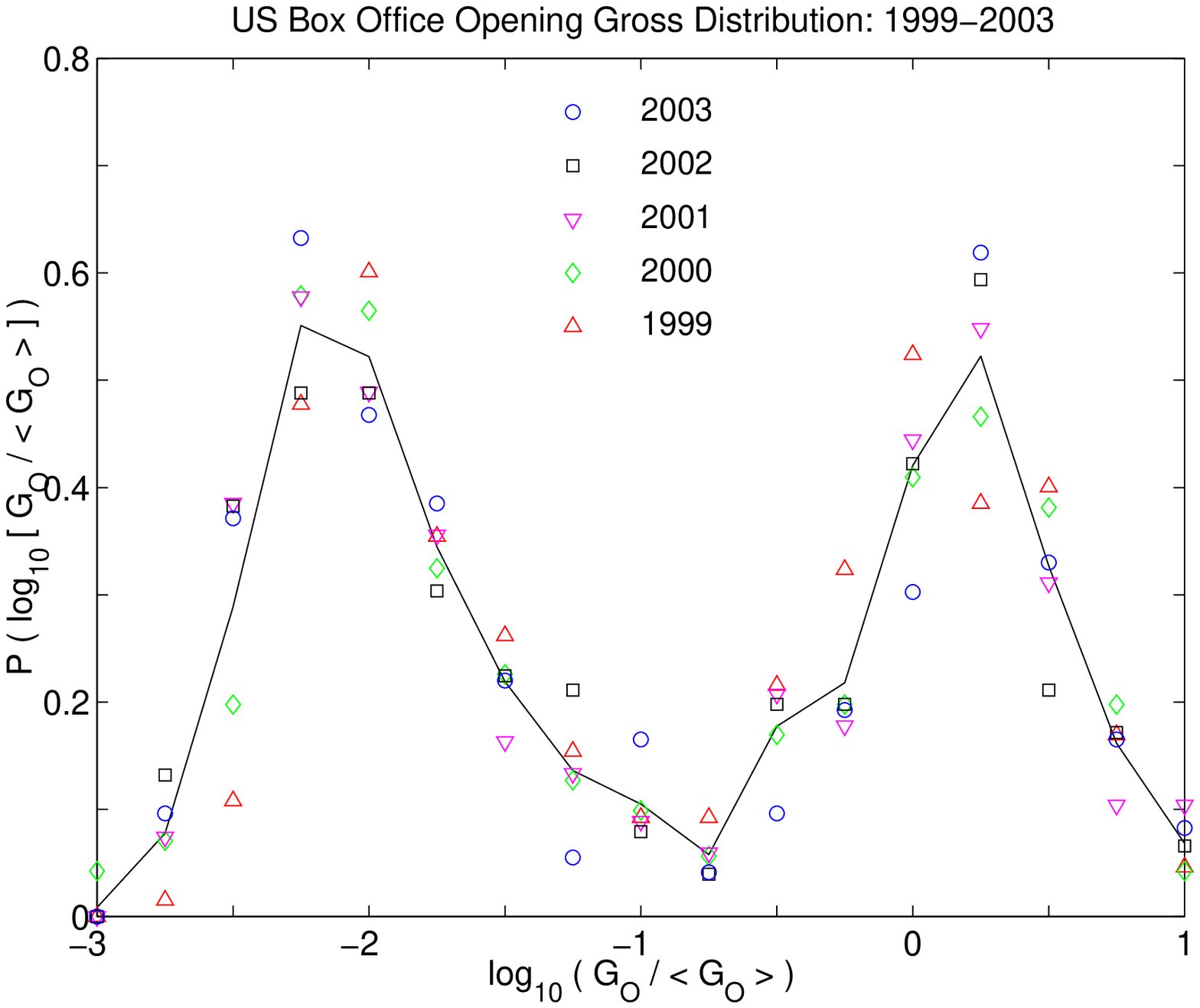}
\includegraphics[width=0.48\linewidth,clip] {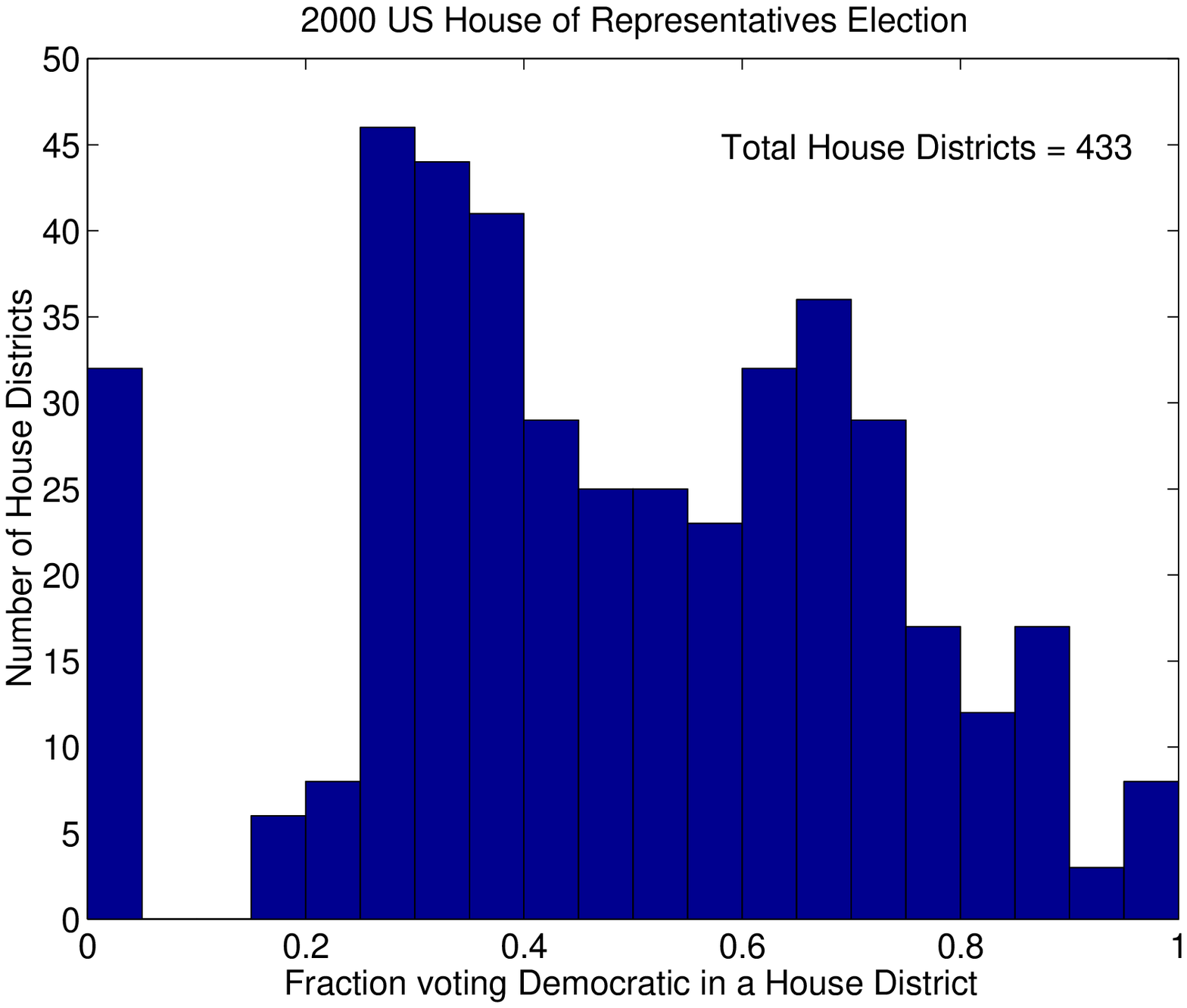}
\caption{Examples of empirical bimodal distributions. (Left) The distribution
of opening week gross earning, $G_O$ (scaled by the average value for a 
particular year, $< G_O >$) for movies released in the USA during the 
period 1999-2003. 
The different symbols
correspond to individual years, while the curve represents the average
over the entire period.
(Right) Frequency histogram of vote share for the Democratic Party candidate
in each House district at the 2000 US Federal House of Representatives 
election.}
\label{fig:a}       
\end{figure}

To account for this we argue that, in a
society, agents make choices based on their personal beliefs
as well as opinions of their neighbors about the possible outcomes of a choice. These beliefs are not fixed but
evolve over time according to changing circumstances, based on previous
choices (adaptation) and how they accorded with those of the 
majority (learning). We
propose a model of collective choice dynamics where each agent has
two variables associated with it, one corresponding to its choice
and the other corresponding to its belief regarding the possible outcome
of the choice. 

The bounded rationality of the agents in our model is due to the limited
information available to the agent at a given point of time. However,
subject to this constraint, the agent behaves deterministically. One of
the striking observations obtained from the model is that although each
agent may behave rationally and change their beliefs (and hence
their choices) periodically, the collective choice 
may get polarized and remain so
for extremely long times (e.g., the entire duration of the simulation).

\vspace{-0.5cm}
\section{The Model}
\vspace{-0.25cm}
Our model is defined as follows.
Consider a population of $N$ agents, each of whom can be in one of
two choice states $S = \pm 1$ (e.g., to buy or to sell,
to vote Party A or Party B, etc.). In addition, each agent has a personal
preference or belief, $\theta$, that is chosen from a uniform random
distribution initially.
At each time step, every agent considers the average choice of its
neighbors at the previous instant, and if this exceeds its belief,
makes the same choice; otherwise, it makes the opposite choice.
Then, for the $i$-th agent, the choice dynamics is described by:
\begin{equation}
S_i^{t+1} = {\rm sign} ( \Sigma_{j \in {\cal N}} J_{ij} S_j^t - \theta_i^t),
\end{equation}
where $\cal N$ is the set of neighbors of agent $i$ ($i = 1, \ldots, N$),
and sign ($z$) = $+1$, if $z > 0$, and = $-1$, otherwise.
The coupling coefficient among agents, $J_{ij}$, is assumed to be a constant
($= 1$) for simplicity and normalized by $z$ ($= |{\cal N}|$), the number of
neighbors. In a lattice, ${\cal N}$ is the set of spatial nearest neighbors and
$z$ is the coordination number.

The individual belief $\theta$ in turn evolves, being incremented or
decreased at each time step, according to the agent's choice:
\begin{equation}
~\theta_i^{t+1} = \theta_i^t + \mu S_i^{t+1} + \lambda S_i^t, ~{\rm if}~
S_i^{t} M^t < 0,
\end{equation}
\vspace{-0.8cm}
$$ =  \theta_i^t + \mu S_i^{t+1}, {\rm otherwise},$$
where $M^t = (1/N) \Sigma_j S_j^t$ is the collective choice
of the entire community at time $t$. The
adaptation parameter $\mu$ is a measure of how frequently an agent switches
from one decision to another. Belief also changes according to whether the
previous choice agreed with the majority decision. In case of disagreement,
the belief is increased/decreased by a quantity $\lambda$ that measures
the relative importance of global feedback (e.g., through information
obtained from the media). 
The desirability of a particular
choice is assumed to be related to the fraction of agents in 
the community choosing
it; hence, at any given time, every agent is trying to coordinate its
choice with that of the majority.

\vspace{-0.5cm}
\section{Results}
\vspace{-0.25cm}
Although some analytical results can be obtained under mean field theory,
here we present only numerical simulation results for the case where the
agents are placed on a two-dimensional 
regular lattice (see Ref. \cite{SFI} for details).
Note that, in absence of either adaptation or global
feedback ($\mu = \lambda = 0$) the model reduces to the well-studied
random field Ising model. 

\begin{figure}[htbp]
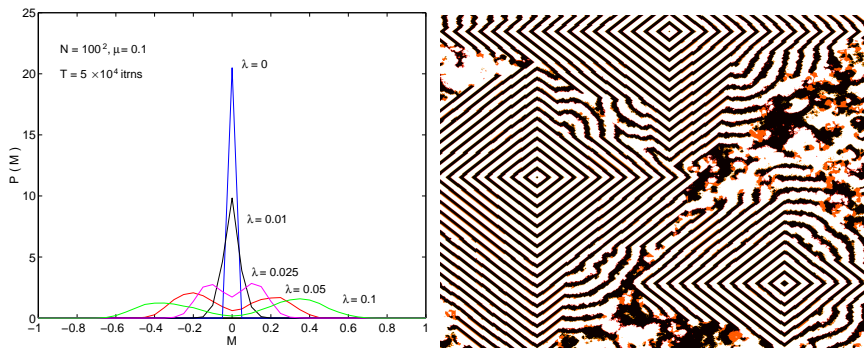

\centering
\includegraphics[width=0.48\linewidth,clip] {fig1.eps}
\includegraphics[width=0.48\linewidth,clip] {fig2.eps}
\caption{(Left) Probability distribution of the collective        
choice $M$ at different values of the global feedback parameter
$\lambda$. A phase transition from a bimodal to an unimodal
distribution occurs as $\lambda \rightarrow 0$. The simulation
results shown are for $100 \times 100$ agents interacting in a 2-D
lattice for 50,000 iterations.
The adaptation rate is $\mu = 0.1$. Compare with Fig. 1a in
Ref. \cite{Ple03}. (Right) Spatial pattern in the
choice behavior for $1000 \times 1000$ agents interacting in
a square lattice after $10^5$ iterations with $\mu = 0.1$ and
$\lambda = 0.05$.
}
\label{fig:1}       
\end{figure}
In the presence of adaptation but absence of learning ($\mu > 0, \lambda = 0$),
starting from an initial
random distribution of choices and personal preferences,
we observe only very small
clusters of similar choice behavior and the average choice $M$ fluctuates
around 0. In other words, at any given time equal number
of agents have opposite choice preferences (on an average). Introduction of
learning in the model ($\lambda > 0$) gives rise to significant clustering
as well as a non-zero value for the collective choice $M$.
We observe a phase transition of the probability distribution of $M$
from an unimodal to a bimodal form as a result of the competition between 
the adaptation and global
feedback effects (Fig. 2 (left)).

The collective choice switches periodically between a positive value and 
a negative value,
having an average residence time which diverges with
$\lambda$ and with $N$.
For instance, when $\lambda$ is very high relative to $\mu$, $M$
gets locked into one of two states
(depending on the initial condition),
corresponding to the majority preferring either one or the other
choice. This is reminiscent of lock-in in certain economic systems subject
to positive feedback \cite{Art89}.
The existence of long-range correlations in
the choice of agents in the bimodal phase often results in striking spatial
patterns such as vortices and spiral waves [Fig. 2 (right)] after long times.
These patterns often show the existence of multiple domains, with the
behavior of agents belonging to a particular domain being highly
correlated and slaved to the choice behavior of an ``opinion leader''.

We have also introduced partial irrationality in the model by making
the choice dynamics stochastic. Each agent may choose the same as
or opposite to that of its neighbors if their overall decision exceeds its
personal belief, according to a certain probability
function with a parameter $\beta$ that is a measure of the degree of
reliability that an agent assigns to the information it receives.
For $\beta \rightarrow 0$, the agent ignores all information and essentially
chooses randomly; in this case, expectedly, the distribution
becomes unimodal. Under mean-field theory, one sees that the bimodal
distribution occurs even for $\lambda = 0$ as $\beta \rightarrow \infty$;
however, as $\beta$ is gradually decreased a phase transition to the unimodal
distribution is observed.

\vspace{-0.5cm}
\section{Discussion and Summary}
\vspace{-0.25cm}
Our model seems to provide an explanation for
the observed bimodality in a large number of social or economic phenomena,
e.g., in the initial reception of movies, as shown
in the distribution of the opening gross of movies released in theaters
across the USA during the period 1997-2003 \cite{Sin04}.
Bimodality in this context implies that movies either achieve
significant success or are dismal box-office failures initially.
We have considered the opening, rather than the total, gross for our
analysis because the former characterizes the uncertainty faced by the
moviegoer (agent) whether to see a newly released movie, when there is
very little information available about its quality. Based
on the model presented here, we conclude that, in such a situation the
moviegoers' choice depends not only on their neighbors' choice about this
movie, but also on how well previous action based on such neighborhood
information agreed with media reports and reviews of movies indicating
the overall or community choice.
Hence, the case of $\lambda > 0$, indicating the reliance of an individual
agent on the aggregate information, imposes correlation among agent choice
across the community which leads to bimodality in the opening gross
distribution.

Our model also provides justification for
the two-phase behavior observed in the financial markets wherein
volume imbalance clearly shows a bimodal distribution beyond a
critical threshold of local noise intensity \cite{Ple03}.
In contrast to many current models, we have not assumed a priori
existence of contrarian and trend-follower strategies among the
agents \cite{Lux95}.
Rather such behavior emerges naturally from the micro-dynamics of
agents' choice behavior. 

Similar behavior possibly underlies biphasic behavior in election results.
The distribution of votes in a two-party election 
will show an unimodal pattern for elections where local issues are
more important than the role of the mass media (hence $\lambda = 0$)
and a bimodal distribution
for elections where voters are more reliant on media coverage
for individual-level voting cues ($\lambda > 0$).

One can also tailor marketing strategies to different segments
of the population depending on the role that global feedback plays in their
decisions. Products, whose consumers have $\lambda = 0$, can be better
disseminated through distributing free samples in neighborhoods; while
for $\lambda > 0$, a mass media campaign blitz will be more effective.

In summary, we have presented here a model of the emergence of collective
choice through interactions between agents who are influenced by their
personal preferences which change over time through processes akin to
adaptation and learning. We find that introducing these effects produce
a two-phase behavior, marked by an unimodal distribution and a bimodal
distribution of the collective choice, respectively. 
Our model explains very well the observed two-phase behavior in 
markets, not only in the restricted context of financial markets, but
also, in a wider context, movie income and election results.
\vspace{-0.4cm}
%
%
\input{referenc}


\printindex
\end{document}

%% file: referenc.tex
%
%

%
%